\documentclass{emulateapj}
\usepackage{aas_macros}
\usepackage{epstopdf}
\usepackage{epstopdf}
\usepackage{epsfig}
\usepackage{natbib}
\usepackage{float}
\usepackage{gensymb}
\usepackage{amsmath}
\usepackage{lipsum}
\usepackage{times}
\usepackage{comment}
\usepackage{color}

\begin{document}

\title{Hydrogen Emission from the Ionized Gaseous Halos of Low Redshift Galaxies}

\author{Huanian Zhang\altaffilmark{1,2}, Dennis Zaritsky\altaffilmark{1}, Guangtun Zhu\altaffilmark{3}, Brice M{\'e}nard\altaffilmark{3} and David W. Hogg\altaffilmark{4}}
\altaffiltext{1}{Steward Observatory, University of Arizona, Tucson, AZ 85719, USA; fantasyzhn@email.arizona.edu}
\altaffiltext{2}{Department of Physics, University of Arizona, Tucson, AZ 85719, USA}
\altaffiltext{3}{Department of Physics \& Astronomy, Johns Hopkins University, 3400 N. Charles Street, Baltimore, MD 21218, USA}
\altaffiltext{4}{Center for Cosmology and Particle Physics, Department of Physics, New York University, 4 Washington Pl., New
York, NY 10003, USA}

\begin{abstract}
Using a sample of nearly half million galaxies, intersected by over 7 million lines of sight from the Sloan Digital Sky Survey Data Release 12, we trace
H$\alpha$ + [N{\small II}] emission from a galactocentric projected radius, $r_p$, of 5 kpc to more than 100 kpc.  The emission flux
surface brightness is $\propto r_p^{-1.9 \pm 0.4}$. We obtain consistent results using only the H$\alpha$ or [N{\small II}] flux. We measure a stronger signal for the
bluer half of the target sample than for the redder half on small scales, $r_p <$ 20 kpc. We obtain a $3\sigma$ detection of H$\alpha$ + [N{\small II}] emission in the 50 to 100 kpc $r_p$ bin. The mean emission flux within this bin is $(1.10 \pm 0.35) \times 10^{-20}$ erg cm$^{-2}$ s$^{-1}$ \AA$^{-1}$, which corresponds to $1.87 \times 10^{-20}$ erg cm$^{-2}$ s$^{-1}$ arcsec$^{-2}$ or 0.0033 Rayleigh. This detection is 34 times fainter than a previous strict limit obtained using deep narrow-band imaging. The faintness of the
signal demonstrates why it has been so difficult to trace recombination radiation out to large radii around galaxies. This signal, combined with published estimates of n$_{\rm H}$, lead us to estimate the temperature of the gas to be 12,000 K, consistent with independent empirical estimates based on metal ion absorption lines and expectations from numerical simulations.
\end{abstract}


\section{Introduction}

A full accounting of the baryons in galaxies is a long standing, but yet unattained goal. Discrepancies between the expected baryon mass, calculated using the cosmological baryon fraction \citep{komatsu}, the total mass of galaxies, and the observed baryonic components (see \cite{miller} for an analysis of the Milky Way's baryon budget that even includes the elusive hot component) remain large.
Reconciliation almost certainly lies with a more complete understanding of the gaseous content within the vast dark matter halos of galaxies, but that component is diffuse and almost certainly in multiple physical phases.

Our theoretical understanding of accretion onto galaxies has evolved, beginning with the initial description of how infalling gas would be shock heated as it entered a dark matter halo \citep{white} to more complex treatments embedded in detailed cosmological simulations \citep[e.g.][]{keres}. A commonality of recent treatments \citep[e.g.][]{nuza,ford} is the resulting complexity of the halo gas and the sensitivity of important resulting details to a variety of poorly constrained factors such as the energetic feedback from the central galaxy and the intensity of the local ionizing radiation field. On the positive side of this complicated interplay is that observations of the baryonic matter in halos have the potential to constrain a variety of phenomenon related to galaxy formation and evolution \citep{rasmussen}.

Interest in tracing the hydrogen gas beyond the inner disks of galaxies, whether it is neutral or ionized, extends well back in time because this material is the presumed reservoir for subsequent star formation \citep{spitzer}. Deep searches for neutral gas in nearby galaxies, based on the 21cm emission feature, hit a floor column density of $\sim 10^{19}$ cm$^{-2}$ \citep{vangorkom}, resulting in apparent sharp edges to the gaseous disks of galaxies plus a limited number of isolated clouds \citep{minchin}. The absence of more widely distributed H{\small I} was attributed to ionization caused  by either the intergalactic UV background  \citep{maloney} or escaping radiation from the galaxy itself \citep{tinsley,heckman}. This hypothesis, in turn, helped motivate searches for recombination radiation from ionized hydrogen in galaxy halos \citep{bland-hawthorn}.

The emission measure of such gas is exceedingly small, demanding either deep long slit spectroscopy \citep{christlein} or integral field spectroscopy \citep[e.g.][]{bland-hawthorn,dicaire,hlav,adams}. The most stringent limits on H$\alpha$ emission in the halos of nearby galaxies placed by the last of those studies is a
5$\sigma$ upper limit of $6.4 \times 10^{-19}$ erg s$^{-1}$ cm$^{-2}$ arcsec$^{-2}$, which informs us of the minimum required sensitivity to make further inroads.

Despite the lack of detection so far, we do expect to eventually observe recombination radiation from diffuse ionized hydrogen in galactic halos. Absorption line spectroscopy of QSOs has long associated metal lines and stronger Ly $\alpha$ absorption lines with galaxies \citep{young,bergeron,rauch,bowen} and studies of our own galaxy have also identified ionized metals \citep{pettini,putman} with a large covering factor \citep[$\sim$ 91\%;][]{collins}. Furthermore, detection of hydrogen emission from individual high velocity clouds in our galaxy do exist \citep[e.g.][]{weiner,bland-hawthorn98,sembach} and such emission is extensive around the Magellanic Stream \citep{fox}. The latter argues for vast amounts of H{\small II} associated with the Magellanic Stream's H{\small I}.

We approach the observational challenge in a similar manner as done previously in searches for metal absorption lines in nearby galaxies, by combining spectra of numerous lines of sight that intersect the halos of nearby galaxies \citep{bothun,cote}. However, those studies required bright background sources against which to search for the absorption. We are searching for emission and therefore prefer to utilize lines of sight with as little contribution from any other source as possible. The great advantage we have over previous studies is the availability of the massive spectroscopic database generated by the SDSS project. Instead of relying on a handful of lines of sight around one to a few target galaxies, we will be combining millions of target/line-of-sight pairs. We describe the selection of the target or primary galaxy sample
and the associated lines of sight in \S \ref{sec:DAM}, tests of our methodology in \S \ref{sec:tests}, our results in \S \ref{sec:ourresults}, including comparisons of the halo properties of red vs. blue galaxies and lower vs. higher luminosity galaxies. We briefly place the results in context, including deriving the temperature of the gas, in \S \ref{sec:discussion}. Finally we summarize our study in \S \ref{sec:summary}. Throughout, we adopt
$\Omega_m$ = 0.3, $\Omega_\Lambda =$ 0.7, $\Omega_k$ = 0 and the dimensionless Hubble constant $h = $ 0.7.


\section{Data and Analysis Methodology}
\label{sec:DAM}

We obtain the data for this study from the Sloan Digital Sky Survey Data Release photometric and spectroscopic catalogs  \cite[SDSS DR12;][]{2015ApJS..219...12A}.
To probe a target galaxy's halo we could use any SDSS spectrum  of a line of sight that is projected sufficiently near that target galaxy. However, there are several reasons for limiting both the sample of primary targets, those galaxies whose halos are probed, and secondary targets, those galaxies that are projected nearby for which SDSS spectra exist. We describe the selection criteria below.
After identifying the suitable SDSS spectra, we obtain the associated wavelength-calibrated, flux-calibrated, and sky-subtracted SDSS spectra that have been rebinned onto a uniform wavelength grid with $\Delta\log_{10}\lambda = 10^{-4}$ \citep{2012AJ....144..144B}. In addition, we use the SDSS photometric measurements (magnitudes, sizes) and object classifications (eg. STAR vs. GALAXY). We use only objects that are classified as a GALAXY by SDSS in all of our subsequent analysis. The object type of the secondary target is in principle irrelevant, but we found QSO and stellar spectra more troublesome to model and subtract, which is why we exclude them.

\subsection{Selecting Primary and Secondary Galaxies}
\label{samples}

To select our sample of primary targets, the galaxies about which we will trace the mean ionized gaseous halo,
we first reject all galaxies that have a redshift, $z< 0.05$. We impose this cut to remove possible misclassified stars and to avoid working at redshifts where there could be Galactic contamination of the H$\alpha$ + [N{\small II}] spectral region. Next, we reject all galaxies with $z > 0.2$. We impose this cut to avoid having redshifted H$\alpha$ + [N{\small II}] land redward of $\sim$ 8000 \AA, where the atmospheric emission is significantly brighter and variable.

\begin{figure}[!htbp]
\begin{center}
\includegraphics[width = 0.50 \textwidth]{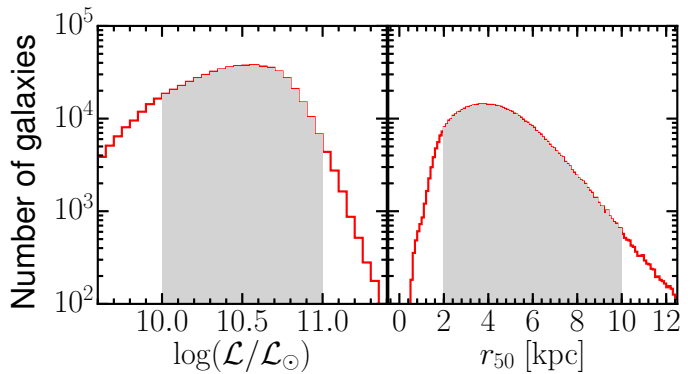}
\end{center}
\caption{The distribution of luminosity and half light radius for candidate primary galaxies. The shaded region marks our selected parameter ranges. Our goal is to ensure a degree of homogeneity in the primary galaxy sample.}
\label{Figure:L_R50}
\end{figure}

Because a concern in any stacking analysis is the degree of homogeneity among the objects contributing to the stack, we also select primaries on the basis of galaxy luminosity, based on ``petroMag\_r", and size, based on the half light radius measurement ``petroR50\_r". The observed parameters are converted to luminosity and physical radius using our adopted cosmological parameters and  $M_{r,\odot} = 4.77$ \citep{2003ApJ...592..819B}. In the interest of limiting heterogeneity in the primary sample we require 10 $< \log \mathcal{L}/\mathcal{L}_\odot \le$ 11
and $2 < r_{50} \le 10$ kpc (as we show in Figure \ref{Figure:L_R50}).

To select our sample of secondary targets, we apply criteria aimed at maximizing the likelihood of a robust measurement (or limit)  of any residual flux at H$\alpha$ + [N{\small II}].
First, however, we search for lines of sight available in the SDSS spectroscopic database that lie near in projection to each primary galaxy.
We calculate the angular distance $\gamma$ on the sky between the two lines of sight using
\begin{equation}
\begin{split}
\cos \gamma & = \cos(90^{\circ} - \alpha_1) \cos(90^{\circ} - \alpha_2) \\
 & + \sin(90^{\circ} - \alpha_1) \sin(90^{\circ} - \alpha_2) \cos(\beta_1 - \beta_2)
\end{split}
\end{equation}
where $\alpha_{1}$ and $\alpha_2$ are the Declination values of the lines of sight (1) and (2) and $\beta_1$  and $\beta_2$ are the corresponding Right Ascension values. We convert to the projected physical separation, $r_p$, using the angular diameter distance \citep{Hogg:1999ad} and retain secondary targets if they have
$r_p \le$ 1.5 Mpc. 
The typical angular distance $\gamma$ between two lines of sight for an accepted pair is $ \sim 0.15^\circ$. 
The distribution of these lines of sight in $r_p$ grows linearly as expected for random pairs for $r_p > 10$ kpc. As mentioned above, we use only lines of sight toward targets classified as GALAXY and apply the $z > 0.05$ to avoid misclassified stars. Unlike for the primary galaxies, we do not impose an upper limit on the redshift of the secondary targets.

After identifying potential line of sight pairs, we trim further by requiring
that the secondary target differ in redshift from the primary target by at least 0.05 to avoid confusing a spectral line in the secondary target with one in the primary target. After applying the constraints discussed so far, we have a sample of 9,611,765 pairs of primary and secondary galaxies.
We will cut further based on continuum brightness, but we describe that criterion below, after describing how we fit the continuum and the resulting stacking of spectra.

\begin{figure}[!htbp]
\begin{center}
\includegraphics[width = 0.36 \textwidth]{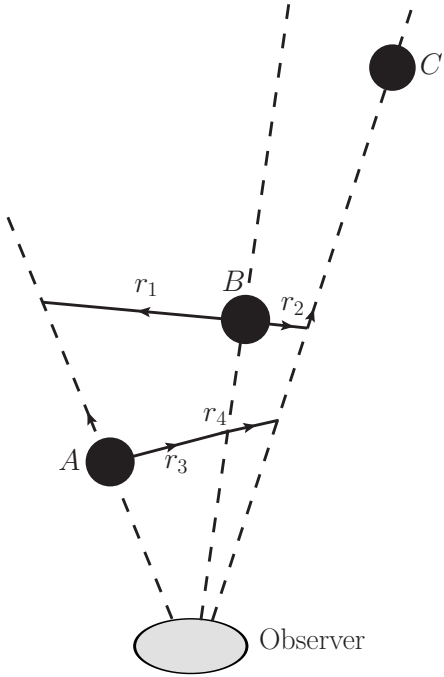}
\end{center}
\caption{Illustration of primary - secondary target geometry. SDSS spectroscopically  targets three galaxies, A, B, and C in this scenario that all differ in redshift by more than 0.05. A and B also happen to have $0.05 < z < 0.2$ and are therefore candidate primary targets for our study. If the projected separations, $r_p$, between the lines of sight to the galaxies, at the redshift of the appropriate primary galaxy is less than 1.5 Mpc, then each of these spectra help probe the halos of galaxies A and B. The spectrum of galaxy A may contain H$\alpha$+[N{\small II}] emission in the halo of B at $r_p = r_1$. The spectrum of galaxy B may contain such emission from the halo of A at $r_p = r_3$. Finally, the spectrum of galaxy C may contain such emission from both A, at $r_p=r_4$, and B, at $r_p = r_2$. Because these are all at sufficiently different redshifts, there is no confusion as to the origin of any emission that is identified.}
\label{Figure:galaxypair}
\end{figure}

To be specific, we illustrate the geometry of primary and secondary lines of sight in Figure \ref{Figure:galaxypair} using a scenario with three galaxies that have SDSS spectroscopic data and are all projected within 1.5 Mpc of each other, two lie at $z < 0.2$ and one at $z > 0.2$. Because of the specified redshifts, only two qualify as primary galaxies but they each provide lines of sight probes of the halos of the other primary galaxies in the example. In total, therefore, these three galaxies provide four lines of sight that probe the halo of the average galaxy at four different projected radii. Note that the galaxy along a secondary line of sight can be at a redshift lower than that of the primary galaxy.

\subsection{Masking and Continuum Fitting}
\label{sec:fitting}

The spectrum of the secondary target is irrelevant and frankly serves only to add noise to our intended goal, the target halo
H$\alpha$ + [N{\small II}] emission. This unwanted contribution
can be divided into a relatively smooth continuum and absorption/emission lines. The former we fit and subtract as described next, the latter we mask and select against as described further below. We estimate the smooth component  in two different ways, either by fitting a tenth order polynomial, the ``polynomial" method, or by applying a 25 pixel median filter, the ``median" method. Using two approaches provides us with an indication of the systematic uncertainties arising from the continuum subtraction methodology, but the polynomial fitting is our preferred  and default approach. Because even the smooth continuum of galaxies can be complex, we limit our analysis to the continuum within the proximity of H$\alpha$ ($6340-6790$ \AA) at the redshift of the primary galaxy.
We then translate the residual spectrum to the primary galaxy's rest frame.

\begin{deluxetable}{ll}
\tablecaption{Spectral Features}
\tablewidth{0pt}
\tablehead{
\colhead{Name} & \colhead{$\lambda_0$ (\AA)} \\
}
\startdata
$[$OII$]$ & 3727 \\
H9 & 3835 \\
CaII K & 3935 \\
CaII H & 3970 \\
H$\delta$ & 4102 \\
G-band & 4308 \\
H$\beta$ & 4861 \\
$[$OIII$]$ & 4959 \\
$[$OIII$]$ & 5007\\
Mg b & 5173 \\
NaI & 5896 \\
$[$SII$]$ & 6716 \\
$[$SII$]$ & 6731 \\
$\oplus$ & 7246 \\
$\oplus$ & 7605 
\enddata
\label{table:backglines}
\end{deluxetable}

To remove cases where we clearly have difficulties fitting the continuum and where the residual noise is large due to the signal from the secondary target, we reject secondary targets if the mean  continuum flux, $\bar C$, over the region of interest is $> 3.0 \times 10^{-17} \ {\rm erg} \ {\rm cm}^{-2} \ {\rm s}^{-1} \ {\rm \AA}^{-1}$. We set this threshold after analyzing the stacked residual spectra that we describe below. In summary, we find that stacks of spectra with continua greater than this level show systematic deviations from a near zero residual mean, presumably because our continuum estimates are an inadequate match for bright complicated continua. Applying this cut leaves us  with 7,567,769 pairs.

In more detail, the continuum estimation, whether via fitting or median filtering, is an iterative scheme where we first mask
a number of spectral features, both in the spectrum of the secondary target and the sky ``background", that can distort the estimation.
The features we address are those in the secondary target that could be shifted by the relative redshift difference between primary and secondary
into the wavelength interval of interest
and those in the background onto which the primary spectrum could be redshifted
(Table \ref{table:backglines}). To reduce the effect of these strong spectral features on the continuum estimation, we mask $\pm$ 18 \AA\ from each of these lines if they lie within the range of wavelengths that we use. We select $\pm$ 18 \AA\ as a range based on a visual examination of the broadest features in the deep background stack that we discuss below (\S \ref{sec:sky}). Additionally, in each of two polynomial fitting iterations we remove 2$\sigma$ outliers from the fitting calculation. 

Once the smooth continuum estimate is complete, we subtract it from the original spectrum and mask $\pm$ 18 \AA\ about each of the strong spectral features in Table \ref{table:backglines} by replacing the residual flux values with zeros. We shift the relevant portion of the original spectrum onto a fixed region in rest frame wavelength, $6415-6715$ \AA. The interpolation onto this new wavelength grid is done simply by populating with the nearest pixel value.
Lastly, we use the mean value of the residual flux over this window, $\bar{R}$, as a diagnostic of potential problems in the continuum subtraction. We reject spectra for which $\bar{R}$ is an outlier in the distribution ($>$ 2$\sigma$). This cut results in rejecting spectra that have $|\bar{R}| > 0.04\times 10^{-17} \ {\rm erg} \ {\rm cm}^{-2} \ {\rm s}^{-1} \ {\rm \AA}^{-1}$.

\begin{figure}[!htbp]
\begin{center}
\includegraphics[width = 0.48 \textwidth]{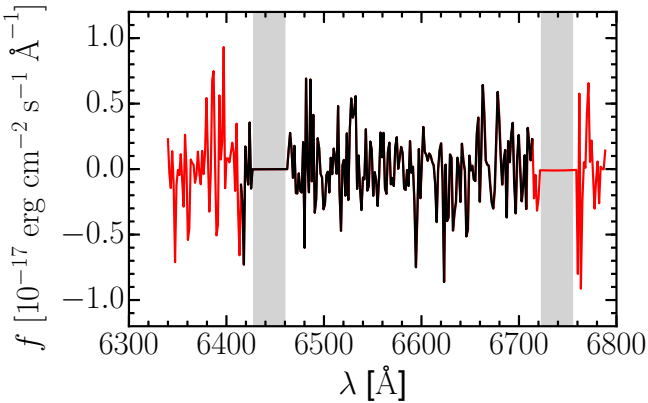}
\end{center}
\caption{We interpolate our continuum-subtracted, de-redshifted spectral segments onto a defined wavelength grid over the spectral region from $6415-6715$ \AA\ (rest frame). The lighter line (red in the color version) shows the region extracted from the original spectrum for continuum subtraction, while the darker overlaid line shows the interpolated spectrum over the fixed grid for a single line of sight. Regions with an $f$ value of zero that are also shaded are masked regions.}
\label{Figure:intplt}
\end{figure}

These criteria define our final sample of spectra and
result in a sample of 7,263,239 lines of sight that intersect the halos of 491,469 galaxies. We summarize the criteria and their impact on reducing the sample size in Table \ref{table:cutflow}. 
These continuum-subtracted, de-redshifted, interpolated spectral segments, as that shown in Figure \ref{Figure:intplt}, are the basis of our subsequent analysis.
\begin{deluxetable}{lc}
\tablewidth{0pt}
\tablecaption{Cut Criteria and Sample Size}
\tablehead{
\colhead{Sample selection\tablenotemark{a}} & \colhead{Fraction of Original Sample} \\
}
\startdata
10 $< \log(\mathcal{L}/\mathcal{L_{\odot}}) \le$ 11 & 73.2\%\\
2 kpc $< r_{50} \le$ 10 kpc & 70.0\%\\
$\bar C \le$ $3.0\times 10^{-17}$ erg cm$^{-2}$ s$^{-1}$ \AA$^{-1}$ & 54.7\% \\
$|\bar{R}| \le 0.04 ~\times 10^{-17}$ erg cm$^{-2}$ s$^{-1}$ \AA$^{-1}$ & 53.5\%
\enddata
\label{table:cutflow}
\tablenotetext{a}{Cuts are applied in series beginning from the first listed.}
\end{deluxetable}


\section{Results}
\label{sec:results}

Once the spectra are continuum subtracted and rebinned onto a uniform rest wavelength grid, we analyze them in two different ways. First, we combine subsets of the data to produce mean, or composite, spectra for lines of sight that meet specific criteria, such as being within a certain $r_p$ range. Second, we measure the H$\alpha$ and/or [N{\small II}] fluxes in individual lines of sight and then analyze the ensemble of measurements. We present results using both approaches. The former is easier to visualize, the latter retains more information content.

\subsection{Tests and Stacked Spectra}
\label{sec:tests}

\begin{figure}
\begin{center}
\includegraphics[width = 0.48 \textwidth]{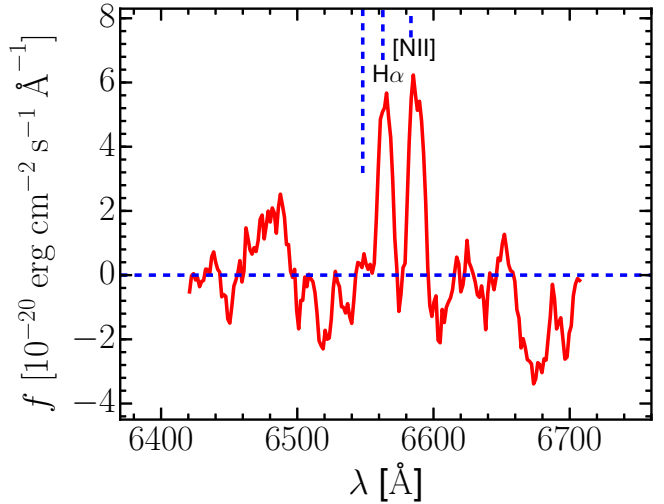}
\end{center}
\caption{The boxcar smoothed composite spectra at $r_p <$ 50 kpc, where the boxcar width corresponds to $\pm$275 km s$^{-1}$ or approximately the maximum rotation velocities of our primary galaxies.  We label the location of H$\alpha$ and the two satellite [NII] lines with vertical dashed lines. The peak values of H$\alpha$ and the redder of the [N II] lines constitute $>4\sigma$ detections and the lines are at the expected wavelengths.}
\label{Figure:stack_signal}
\end{figure}

We present the boxcar smoothed composite spectra in the radius bin $r_p <$ 50 kpc in Figure \ref{Figure:stack_signal}. The peak value of H$\alpha$ and [N{\small II}]6583 ($\sim 6.0 \pm 1.5$ in unit of 10$^{-20}$ erg cm$^{-2}$ s$^{-1}$ \AA$^{-1}$) constitute a 4$\sigma$ detection.
In the following, we
present a set of tests of the procedure that also enable us to understand our sensitivity limits and limiting noise characteristics.

\subsubsection{The Background Test}
\label{sec:sky}

The original, unprocessed SDSS spectra are a combination of ``backgrounds" or ``sky", such as atmospheric continuum, emission lines, and absorption lines, and ``source". The SDSS pipeline implements an extensive effort to provide source spectra from which the sky has been removed. By stacking spectra, we examine whether there are any systematic problems with the sky determination and the level of uncertainty that results either from remaining sky contamination or simply from the increased noise in regions of higher backgrounds.

\begin{figure}[!htbp]
\begin{center}
\includegraphics[width = 0.48 \textwidth]{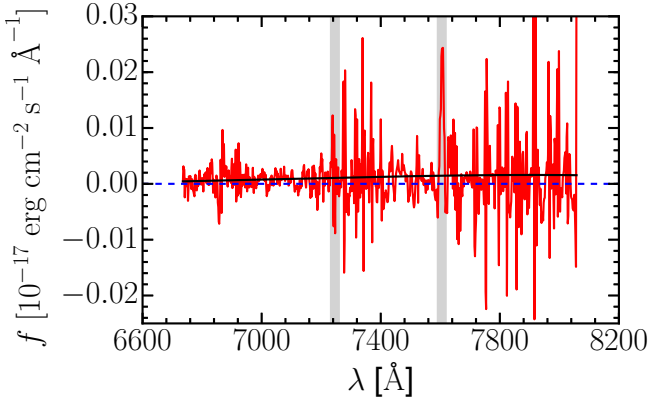}
\end{center}
\caption{Average residual sky background from a stack of 7,245,797 spectra. The solid curve is the best fit polynomial to the sky background. The shaded regions coincide with strong atmospheric absorption features that are masked in our standard analysis but not in this stack. The remaining ``noise"  comes from residuals from the subtraction of strong atmospheric emission features that increase in number and strength toward redder wavelengths. The maximum residuals are $\sim 0.03\times 10^{-17}$ erg cm$^{-2}$ s$^{-1}$ \AA$^{-1}$.}
\label{Figure:skybackg}
\end{figure}

We apply the full procedure as described above, except that rather than shifting the spectra into the primary galaxy rest frame and analyzing the region around H$\alpha$, we keep the spectra in the observed frame. In addition, we do not mask the two atmospheric absorption features listed in Table \ref{table:backglines} and do not trim to the region around H$\alpha$. We stack 7,245,797 spectra, which
constructively combines features that are fixed in the observed frame.
We show the result of this exercise, the mean residual spectrum in the observed frame, in Figure \ref{Figure:skybackg}.

There are various results of note here. First, the residual stacked spectrum is dominated by noise. This demonstrates that the SDSS sky subtraction is not introducing systematics into our analysis. Second, the noise increases redward as expected and even more beyond our imposed wavelength cut (not shown in Figure). Third, our two highlighted atmospheric absorption bands are well treated on average, but we retain the masking in our subsequent analysis because we have noticed these features remain significant in some individual spectra. Fourth,
there may be hints of additional spectral features that appear to be real sky features, such as the absorption at $\sim$ 6850 \AA, but mostly what is visible are noise residuals.  Because any real features or systematic errors that are visible in this Figure will appear at different primary target, rest frame wavelengths, the residuals act as an additional source of noise but will be incoherent.  The fractional noise they introduce will decrease with sample size. The rms of the residual background contribution rises from $\sim$ 0.002 at the bluer end to $>$ 0.01 in units of $10^{-17}$ erg cm$^{-2}$ s$^{-1}$ \AA$^{-1}$.
Even for the larger rms values, $\sim 0.03$ in these units, when we combine 1,000 spectra
the uncertainty due to background residuals will be $\sim 10^{-20}$ erg cm$^{-2}$ s$^{-1}$ \AA$^{-1}$, assuming the sky residuals are randomly distributed in target galaxy, rest frame wavelength. We conclude, given that we always analyze well above 1,000 systems, that this is not our
dominant source of uncertainty. Fifth, there is a hint from our polynomial fitting that the mean value of the residual is positive. We will discuss systematic deviations from zero mean below.

We have just introduced a measurement of our potential sensitivity, $\sim 10^{-20}$ erg cm$^{-2}$ s$^{-1}$ \AA$^{-1}$, so we need to explain how to compare this value to what is 
common in the literature.
Recall that current limits on ionized flux in galactic halos are $\sim 6.4 \times 10^{-19}$ erg s$^{-1}$ cm$^{-2}$ arcsec$^{-2}$ \citep{adams}. There is a slight mismatch in units between the measurements we quote and those typically quoted in the literature because we, by construction, integrate over the fiber aperture and narrow-band imaging studies, by construction, integrate over the wavelength width of the filter. To translate between the
two sets of measurements, consider that our emission lines are generally $\sim$ 12 \AA\ full width ($\sim$ 550 km sec$^{-1}$) and the SDSS fibers are 3 arcsec in diameter (angular area $\sim 7$ arcsec$^{2}$). The rough conversion between our units and those in the literature therefore requires multiplying our values by 1.7. This conversion can be wrong if the emission is either highly concentrated spatially or in velocity. Nevertheless, sensitivity limits at the level of
10$^{-20}$ erg cm$^{-2}$ s$^{-1}$ \AA$^{-1}$, in our units, would constitute a significant improvement over previous limits.

\subsubsection{The Stacked Secondary Targets Test}

\begin{figure*}
\begin{center}
\includegraphics[width = 0.8 \textwidth]{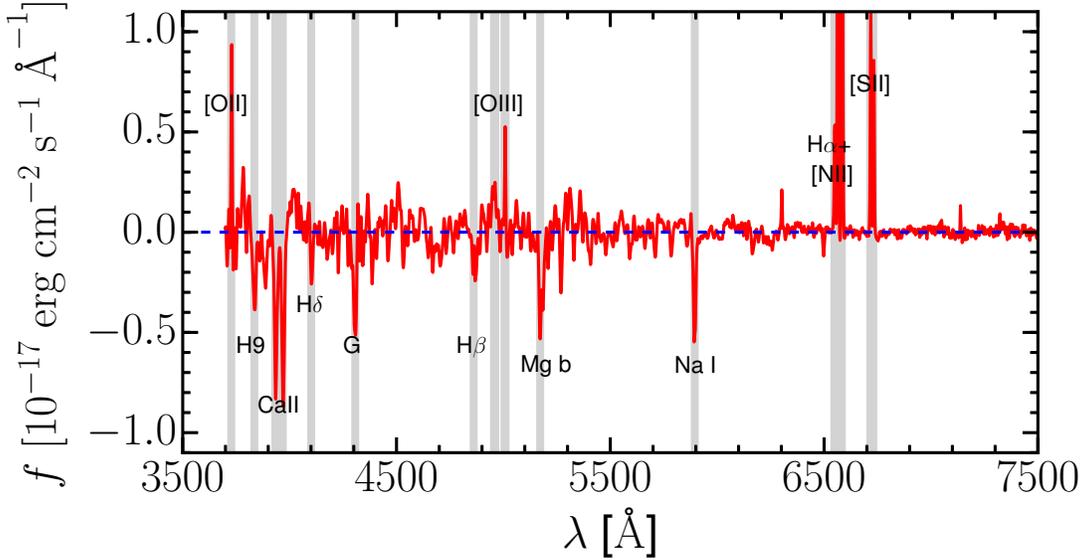}
\end{center}
\caption{Mean, continuum subtracted spectrum of secondary targets. The grey shaded regions indicate the masked regions around well known spectral features (Table \ref{table:backglines}).}
\label{Figure:zbackg}
\end{figure*}

To test our procedure and confirm that we can coherently combine spectra, we now reapply our method,  but this time we shift the spectra using the redshifts of the secondary galaxies.
As such, the stack will be that of an average secondary galaxy spectrum.
This exercise allows us to test the method in a different way and to see what other spectral features may be present that we did not include
in Table \ref{table:backglines}. Here we use a wider spectral region than in our primary galaxy analysis, examining rest frame wavelengths of $3700 - 7500$ \AA. Any portion of the spectrum that happens to lie beyond an observed wavelength of 8500 \AA\ we replace with zeros to avoid introducing large background residuals into the stack. We stack 4,385,225 spectra and show the
results in Figure \ref{Figure:zbackg}.

The major spectral lines, those in Table \ref{table:backglines}, are clearly visible and highlighted. In addition there are now a variety of other features that can be
seen. Again, because these will appear nearly randomly distributed when we shift to the primary galaxy rest frame, these can be considered to be an additional source of noise. Here, however, the noise can be significantly larger than what we found for the sky residuals. Some unmasked features reach $\pm 0.3 \times 10^{-17}$ erg cm$^{-2}$ s$^{-1}$ \AA$^{-1}$, a factor of 10 larger than for the sky residuals. As such we need to combine a factor of 100 more spectra to reach the equivalent level of noise, but even 100,000 lines of sight is not beyond the range of what we can do with our sample, demonstrating that we will be able to reach limits
below 10$^{-20}$ erg cm$^{-2}$ s$^{-1}$ \AA$^{-1}$ in sensitivity, but perhaps not much better than that level with the current sample. For this reason, we must mask bright lines in the secondary target.

\subsubsection{Random Redshift Test}

\begin{figure*}
\begin{center}
\includegraphics[width = 0.8 \textwidth]{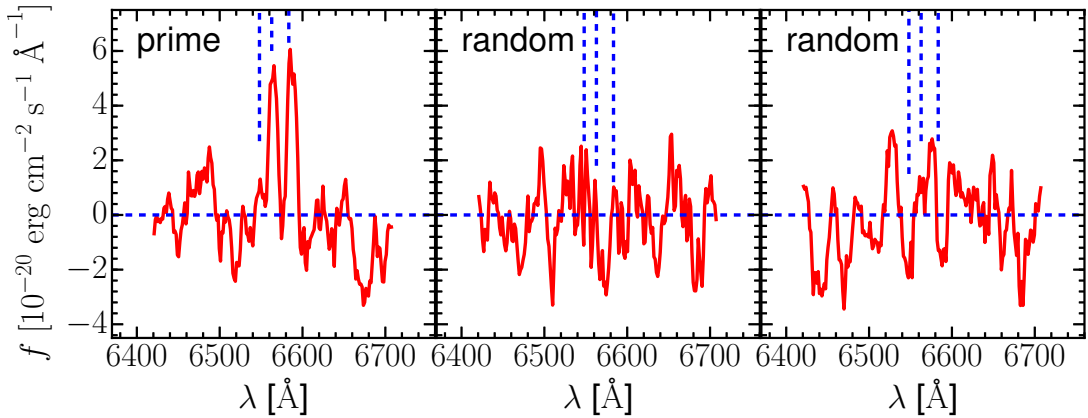}
\end{center}
\caption{The random redshift test. In each panel we present boxcar smoothed composite spectra, where the boxcar width corresponds to $\pm$275 km s$^{-1}$ or approximately the maximum rotation velocities of our primary galaxies.  We label the location of H$\alpha$ and the two [NII] lines with the vertical dashed lines. The leftmost panel presents the stack using the actual primary target redshifts and shows evidence for H$\alpha$ and [NII]6583. The peak values of these lines constitute $>4\sigma$ detections and are at the expected wavelengths. The other two panels show two random realizations and no features throughout the wavelength region have peak values that constitute $>3\sigma$ detections. All of these are for the lines of sight at $r_p <$ 50 kpc.}
\label{Figure:stack_rZ}
\end{figure*}

To further test the expected sensitivity, we perform two different tests, the first of which is a test for false positives and is described here. We assign each primary galaxy a random redshift in the range of 0.05 $< z < $ 0.20. We require the difference between the randomly assigned redshift and the true redshift to be larger than 0.05 to avoid a situation where one of the H$\alpha$ + [N{\small II}] lines can pose as another.  We do not expect a signal at H$\alpha$ + [N{\small II}] due to the randomized redshifts and this is confirmed in the stacked spectra from two realizations that we compare in Figure \ref{Figure:stack_rZ} to the stack constructed using the actual redshifts (reprised here from Figure \ref{Figure:stack_signal}).
We confine our test to lines of sight with $r_p < 50$ kpc, where the true stack shows the strongest visible emission to provide clearest contrast with the random redshift stacks. 

The random redshift stacks provide a guide to the level of uncertainties. For example, excluding H$\alpha$ and [N{\small II}], the spectra in Figure \ref{Figure:stack_rZ} have candidate absorption or emission features with peak $|f| < 3.5 \times 10^{-20}$ erg cm$^{-2}$ s$^{-1}$ \AA$^{-1}$. The standard deviation of the random redshift spectra is $1.5 \times 10^{-20}$ erg cm$^{-2}$ s$^{-1}$ \AA$^{-1}$, demonstrating that none of the candidate features in the random redshift spectra are significant ($> 3 \sigma$). In contrast,  H$\alpha$ and [NII] 6583\AA $~$ with a peak flux of $\sim 6 \times 10^{-20}$ erg cm$^{-2}$ s$^{-1}$ \AA$^{-1}$ are significant, and even more so because they are features found at the anticipated wavelengths.
In the analysis described below, we utilize the full ensemble of measurements from individual lines of sight rather than these stacks. We will reprise the random redshift test in that context below.

\subsubsection{Line Injection Test}

\begin{figure}
\begin{center}
\includegraphics[width = 0.48 \textwidth]{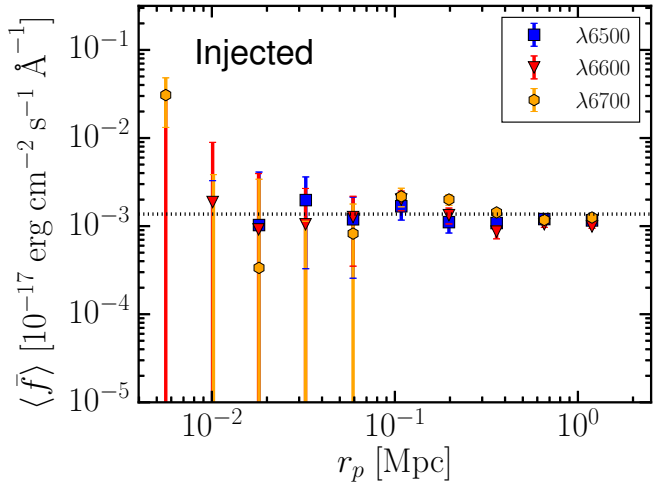}
\end{center}
\caption{Results from the line injection test. We present the measurements within the injected line window. The injected signal is recovered independent of wavelength or radius for $r_p > 100$ kpc, at a level consistent with the input. Uncertainties become sufficiently small for this level of signal once the number of spectra contributing to the stack is sufficiently large. $\langle \bar f \rangle$ is defined in detail in Sec. \ref{sec:ourresults}. The horizontal line represents the level of the injected signal.}
\label{Figure:line_injection}
\end{figure}

The second of our sensitivity tests involves injecting a line of known flux into the individual spectra and reproducing our measurements. 
We inject a Gaussian signal with peak amplitude $2.0\times 10^{-20} \ {\rm erg} \ {\rm cm}^{-2} \ {\rm s}^{-1} \ {\rm \AA}^{-1}$ with width $\pm$275 km s$^{-1}$ (corresponding to an average flux of $1.37\times 10^{-20} \ {\rm erg} \ {\rm cm}^{-2} \ {\rm s}^{-1} \ {\rm \AA}^{-1}$ in this window).
The strength of the injected signal is set to match what we are detecting in our 50 to 100 kpc bin
(see \S\ref{sec:ourresults}). 
To examine our sensitivity as a function of where in the spectrum the line lands, we run the entire
test over for three central injection wavelengths (6500, 6600, and 6700 \AA). 
Although the signal is completely undetectable in a single spectrum, its presence becomes evident once the number of spectra contributing to the stack is sufficiently large to diminish the noise sufficiently. The injected signal is recovered as shown in Figure \ref{Figure:line_injection} independent of wavelength or radius, for $r_p \gtrsim 100$ kpc, at a level consistent with the input. 
We conclude that our limiting sensitivity as determined from injected emission lines is consistent with our detection in the $50 < r_p \le 100$ kpc bin,  $\sim 10^{-20}  \ {\rm erg} \ {\rm cm}^{-2} \ {\rm s}^{-1} \ {\rm \AA}^{-1}$, and is correspondingly higher and lower at smaller and larger $r_p$, respectively. 

\subsection{Results from Individual Line of Sight Measurements}
\label{sec:ourresults}

As an alternative to stacking the spectra, we measure the flux in each individual spectrum coincident with  the emission lines
H$\alpha$ 6562.48\AA, [N{\small II}] 6548.05 and 6583.47\AA\ over a wavelength range corresponding to width $\pm$275 km s$^{-1}$, a value somewhat larger than the characteristic internal
velocities for the type of primary galaxy selected.
We sum the flux in each of the three windows, and then divide by the size of the window to present the mean flux level in the spectral region, $\bar f$.
We reject as outliers those with a mean emission or absorption signal greater than $0.3\times 10^{-17} \ {\rm erg} \ {\rm cm}^{-2} \ {\rm s}^{-1} \ {\rm \AA}^{-1}$.
This cut is intended to eliminate cases where we happened onto a satellite or nearby galaxy with emission or absorption. Note that this level of  emission is below that observed at the edges of disk galaxies \citep{christlein}. This cut affects very few lines of sight (about 0.04\%) and we are left with
7,235,804 lines of sight.

To analyze the results, we group the data in bins of equal $\Delta \log r_p$. This provides a radial sampling weighted to smaller $r_p$ where we anticipate detecting more signal. We calculate a mean value of $\bar f$ in each bin, $\langle \bar f \rangle$, and assign an uncertainty using the internal dispersion and calculating the dispersion in the mean.
We fit to the distribution of
$\langle \bar f \rangle$ vs. $r_p$ using
a power law fitting function plus a constant background term:
\begin{equation}
\label{eq:fitfunc}
\langle \bar f \rangle = a (r_p/50 {\rm\  kpc})^b + c
\end{equation}
where $a$, $b$ and $c$ are fitting parameters.
The parameter $c$ allows for a bias in the determined background level that we first mentioned in \S\ref{sec:sky}. We will discuss this further below, but $c$ is typically non-zero given that there are always uncertainties in the background determinations. We define a new quantity, the net flux, 
$\Delta f \equiv \langle \bar f \rangle - c$.
In our Figures, we also present the median of $\bar s$ corrected for the corresponding $c$ in each radial bin, so that the reader can judge whether the means are unduly influenced by systems with large deviations. We estimate the uncertainty in the median within each bin by calculating the median of the absolute deviations from the median.

\begin{figure*}[!htbp]
\begin{center}
\includegraphics[width = 0.8 \textwidth]{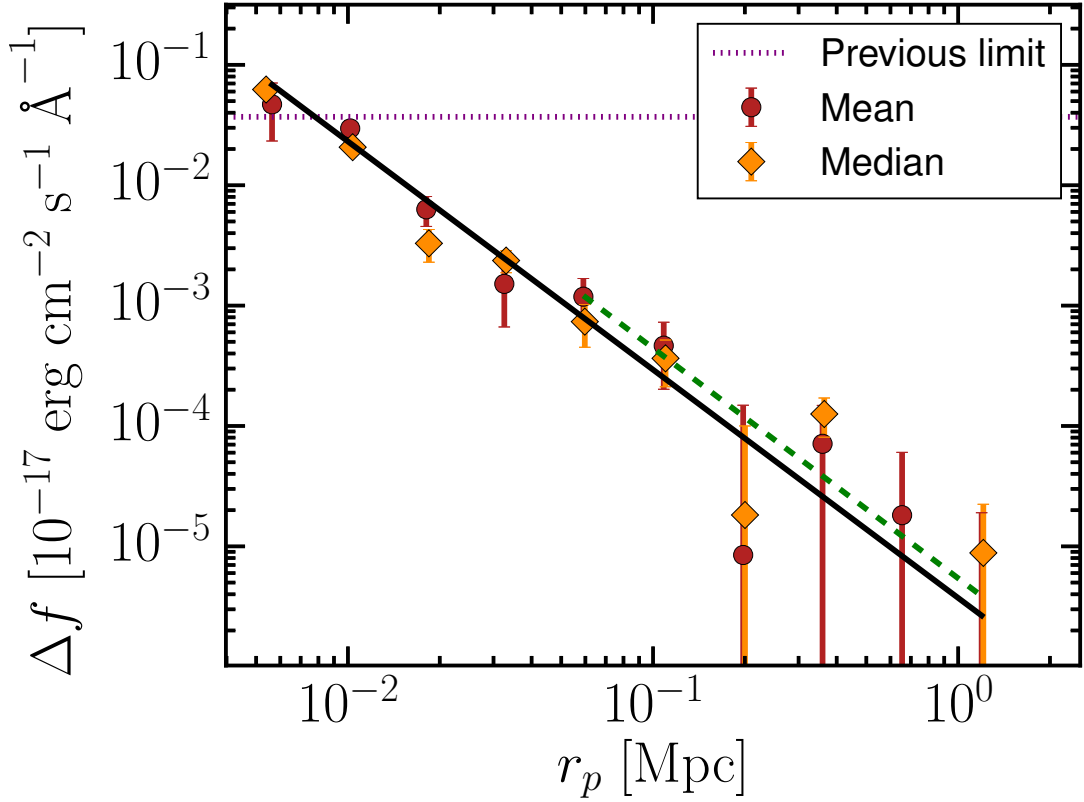}
\end{center}
\caption{The radial profile of  H$\alpha$ +  [NII] around our primary galaxy targets for our entire sample for the polynomial continuum estimator. 
The red filled circles represent the mean values of $\bar f$ in each bin and the orange diamonds the corresponding medians.
We plot two fits, the red solid line corresponds to that which used all of the means, while the black dashed line only to that using the means at $r_p > 50$ kpc. The dotted purple line 
represents the previous limit on H$\alpha$ emission flux from deep narrow-band imaging.}
\label{Figure:logFit}
\end{figure*}

In Figure \ref{Figure:logFit} we present the binned data and our best fit model, Equation \ref{eq:fitfunc}, for our entire sample when we 
adopt the polynomial continuum fitting method.
The median filter continuum estimator shows the same behavior. There is an evident rise in H$\alpha$ + [N{\small II}] toward smaller $r_p$, suggesting a detection of emission from ionized gas in the halos of the galaxies in our sample.
Before proceeding however, there are several aspects of this result that require further discussion. First, the asymptotic level of the data at large $r_p$, the parameter $c$ in our fits, is $> 0$ when we estimate the continua using polynomials and $< 0$ when we use the median filter. We interpret
this difference as indicative of systematic uncertainty in our ability to fully account for the continua ($\sim \pm 10^{-21}$ erg s$^{-1}$ cm$^{-2}$ ${\rm \AA}^{-1}$) . To further explore this issue we return to the random redshift simulations and use them for a third estimate of the background level. The mean measured flux in the two random redshift realizations is $1.50 \times 10^{-21}$ erg s$^{-1}$ cm$^{-2}$ ${\rm \AA}^{-1}$ and $1.27 \times 10^{-21}$ erg s$^{-1}$ cm$^{-2}$ ${\rm \AA}^{-1}$. If we were to assume a background level of zero, then we would have a systematic
uncertainty of the order of $10^{-21}$ erg s$^{-1}$ cm$^{-2}$ ${\rm \AA}^{-1}$. However, because we fit and subtract this bias, our uncertainty is determined by how well we can determine the bias. From Table \ref{table:powerlaw}, we conclude that the uncertainty in $c$ is generally $< 
3 \times 10^{-22}$ erg s$^{-1}$ cm$^{-2}$ ${\rm \AA}^{-1}$. We propagate this uncertainty on the systematic bias into the plotted values of 
$\Delta f$.

\begin{deluxetable*}{llcrrrr}
\tablewidth{0pt}
\tablecaption{Power Law Parameter Fits to $\langle \bar f(r_p) \rangle$}
\tablehead{
\colhead{Primary}&\colhead{Emission}&\colhead{$r_p$ range}& \colhead{a} &\colhead{b} &\colhead{c} & \colhead{$\chi^2$}\\
\colhead{Sample}&\colhead{Lines}&\colhead{[Mpc]}&\colhead{$[\times 10^{-3}]$}&&\colhead{$[\times 10^{-5}]$}} &
\startdata
All& H$\alpha $ + [N{II}] & $r_p < 1.5$ & $1.09 \pm 0.27$ & $-1.90 \pm 0.19$ &$6.2 \pm 2.0$ & 6.79 \\
All& H$\alpha $ + [N{II}] & $0.05 < r_p < 1.5$ & $1.67 \pm 0.46$ & $-1.91 \pm 0.47$ &$5.8 \pm 1.5$ & 2.45\\
All &H$\alpha$ &$r_p < 1.5$ & $0.91 \pm 0.64$ & $-1.91 \pm 0.45$ & $-7.0 \pm 4.7$ & 10.15\\
All &[N II] &$r_p < 1.5$ & $1.18 \pm 0.36$ & $-1.63 \pm 0.24$ & $9.5 \pm 2.7$ & 6.58\\
Red &H$\alpha $ + [N{II}] & $r_p < 1.5$ & $0.76 \pm 0.38$ & $-1.26 \pm 0.40$ & $-2.8\pm 4.7$ & 6.02\\
Blue &H$\alpha $ + [N{II}] & $r_p < 1.5$ & $1.58 \pm 0.55$ & $-1.84 \pm 0.23$ & $ 7.3 \pm 3.9$ & 13.51 \\
Bright &H$\alpha $ + [N{II}] & $r_p < 1.5$ & $1.23 \pm 0.30$ & $-1.72 \pm 0.25$ & $6.8\pm 2.2$ & 2.52\\
Faint &H$\alpha $ + [N{II}] & $r_p < 1.5$ & $1.03 \pm 0.36$ & $-1.95 \pm 0.23$ & $5.4 \pm 2.6$ & 7.07\\
Big & H$\alpha $ + [N{II}] & $r_p < 1.5$ & $0.97 \pm 0.37$ & $-2.35 \pm 0.30$ & $11.4 \pm 2.4$ & 4.34\\
Small & H$\alpha $ + [N{II}] & $r_p < 1.5$ & $0.81 \pm 0.29$ & $-1.90 \pm 0.25$ & $1.7 \pm 2.1$ & 3.99\\
High $z$ & H$\alpha $ + [N{II}] & $r_p < 1.5$ & $1.01 \pm 0.42$ & $-2.06 \pm 0.34$ & $11.7 \pm 2.7$ & 4.62\\
Low $z$ & H$\alpha $ + [N{II}] & $r_p < 1.5$ & $1.12 \pm 0.47$ & $-1.82 \pm 0.30$ & $1.0 \pm 3.3$ & 10.99
\enddata
\label{table:powerlaw}
\tablenotetext{a}{The units for $a$ and $c$ are  $10^{-17} \ {\rm erg} \ {\rm cm}^{-2} \ {\rm s}^{-1} \ {\rm \AA}^{-1}$. The $\chi^2$ for which the best fit is statistically unlikely at 90\% confidence is 12.02 for the 7 degrees of freedom we have. Therefore, our models and the best fit parameters are statistically acceptable across the sample.}
\end{deluxetable*}

The uncertainty in the continuum determination ultimately limits how far in $r_p$ we can trace the emission, but it makes little difference to our primary result --- the detection of emission out to 100 kpc. The value of $c$ changes from
$6.2 \times 10^{-22}$ erg s$^{-1}$ cm$^{-2}$ ${\rm \AA}^{-1}$ to $-8.0 \times 10^{-22}$ erg s$^{-1}$ cm$^{-2}$ ${\rm \AA}^{-1}$ between our two continuum estimation approaches, but the overall fit for the power law behavior is indistinguishable (slopes are 
$-1.90 \pm 0.19$ and $-1.88 \pm 0.17$ and normalizations are $1.09 \pm 0.27$ and $1.13\pm0.27$ for polynomial and median continuum estimators respectively). Significant differences among the data from the two approaches appears only beyond $r_p$ = 100 kpc.

\begin{figure*}[!htbp]
\begin{center}
\includegraphics[width = 0.8 \textwidth]{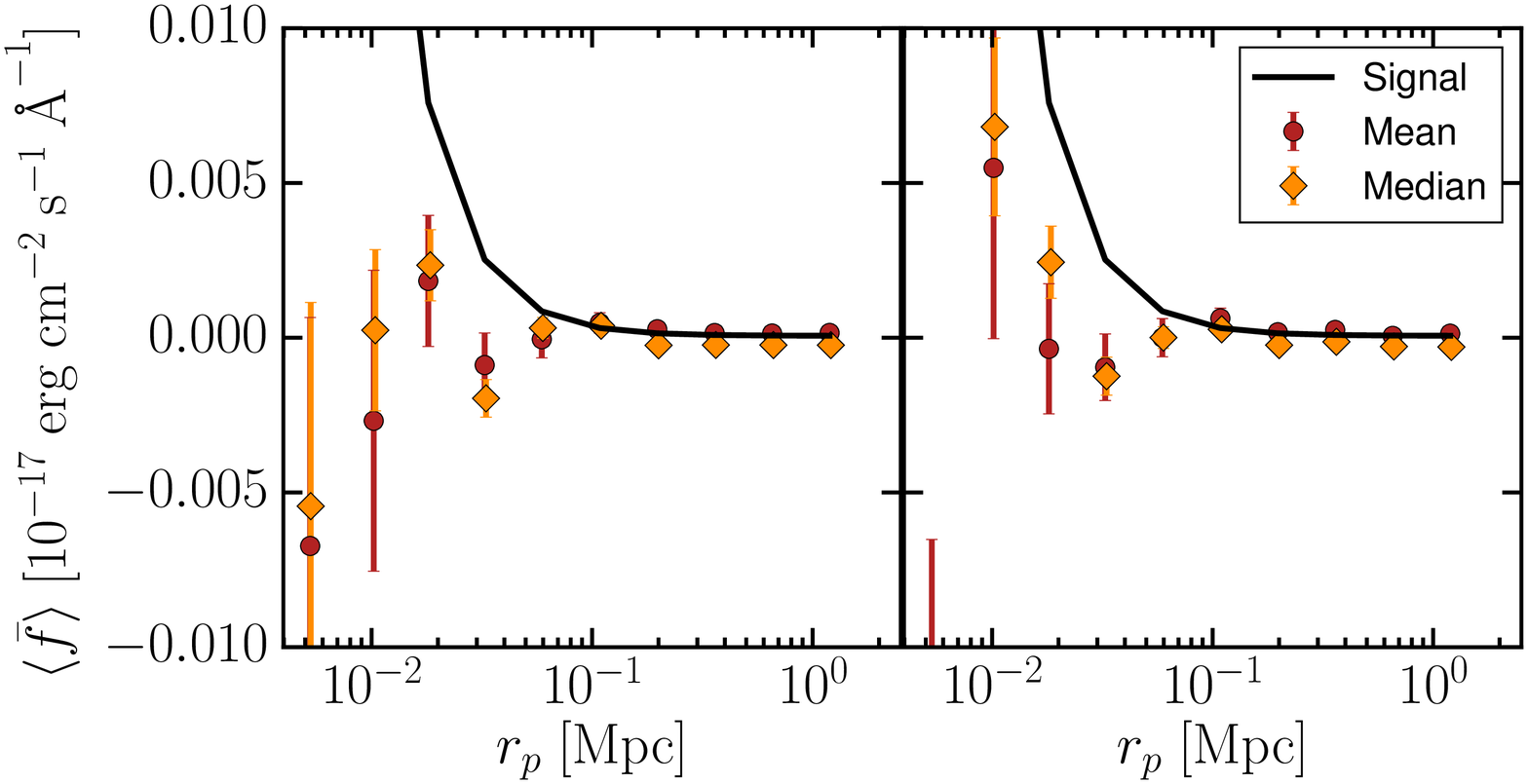}
\end{center}
\caption{The radial profile of H$\alpha$ + [NII] for two samples with randomly shuffled primary galaxy redshifts. 
We plot the means (red circles) and medians (orange diamonds) of $\bar{f}$ for each realization in separate panels.
Both random realizations show a resulting null detection of H$\alpha$ + [NII] over the entire $r_p$ range. 
We also present the best fit power law model of our signal (solid black line) for comparison. Significant detections are limited to $r_p < 100$ kpc.}
\label{Figure:logFit_randomZ}
\end{figure*}

Second, because the uncertainties in $\langle {\bar f} \rangle$ are larger at small $r_p$, due to the smaller number of contributing lines of sight, we investigate whether the apparent rise could be the result of larger random fluctuations at small radii. To address this issue we return to our random redshift simulations.
We present two realization of the random redshift test in Figure \ref{Figure:logFit_randomZ} and both exhibit a null detection of H$\alpha$ + [N{\small II}] in the entire $r_p$ range as seen in the data.
In contrast to the actual fits (Table \ref{table:powerlaw}), where the both the normalization $a$ is $> 0$ and power law index, $b$, is $< 0$ with greater than 2$\sigma$ confidence, in these random redshift tests neither is different than zero with greater than $1 \sigma$ confidence. Unfortunately, executing these random realizations is computationally expensive because they require us to reprocess all of the data. However, we use the measured dispersion in these two simulations to mock additional ones. By drawing randomly from the uncertainties shown in Figure \ref{Figure:logFit_randomZ}, we create 200,000 simulations of the random redshift test and find that only 365 produce $>2 \sigma$ results for positive $a$ and negative $b$ (0.2\%) We conclude that our result is unlikely to be caused by random fluctuations in $\langle \bar f \rangle$. Subsequent tests using subsamples (described below) are consistent and so also argue against a significant role for random fluctuations.

Third, our data extend inward to quite small radii ($<$ 10 kpc). The lines of sights that contribute in this regime do not intersect regions of strong H$\alpha$ emission, because we impose a cut on the line flux, but they should intersect regions near the disk-halo boundary, which we know can have significant emission \citep{rand96}. As such, it is reasonable to suspect that these do not correspond to the more diffuse halo emission that is what we are truly seeking to measure or constrain. To what degree are our fits driven by significant detections at these small impact parameters that may not be truly reflective of a halo distribution?

To address this question, we present the fit using only  data at $r_p > 50$ kpc in Figure \ref{Figure:logFit} and in Table \ref{table:powerlaw}. The resulting fit is consistent with that using all of the data, and still provides statistically significant evidence for an excess of emission in the halo.
We conclude that our results are not begin distorted by emission near the central galaxy.

Fourth, the unprecedented sensitivity of these measurements may mean that we are susceptible to systematic effects that were previously not considered. The one effect that we now consider is scattered light. If the central galaxy is a strong H$\alpha$ emitter, could light scattered either by the Earth's atmosphere or telescope optics, be the source of our signal?

\begin{figure}[!htbp]
\begin{center}
\includegraphics[width = 0.48 \textwidth]{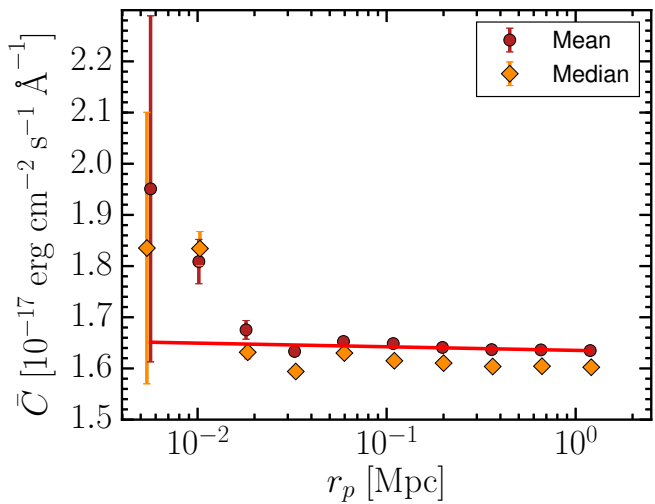}
\end{center}
\caption{The mean continuum flux,  $\bar{C}$, as a function of projected galactocentric radius. We show the means (red circles) and medians (orange diamonds) of $\bar{C}$ in each bin for the polynomial continuum estimator. There is no radial dependence of the continuum beyond $r_p = 10$ kpc. The slope of the fit to the averages is $-0.0020 \pm 0.0006$.}
\label{fig:scatter_light}
\end{figure}

To address this issue, we examine whether continuum light is scattered into our apertures. Our background estimates are dominated by light from the secondary galaxy, but would also include scattered primary galaxy light if such scattering is significant. We plot in Figure \ref{fig:scatter_light} the mean and median values of the mean continuum flux around H$\alpha$, $\bar{C}$, as a function of $r_p$. We detect flux from the primary galaxies at $r_p \sim $ 10 kpc ($\sim 5.5$ arcsec for a typical primary galaxy redshift of 0.1), but beyond this radius there is no statistically significant radial dependence, demonstrating that light from inner radii is not being scattered to larger radii. 
We also compare the results obtained using the nearer and farther halves of the primaries. If the detected signal is due to scattered light then the emission halos of the more distant galaxies will appear to be physically larger. The ratio of scale lengths (see Table \ref{table:powerlaw}) is
$1.13 \pm 0.26$, consistent with a ratio of 1 and inconsistent at the 2$\sigma$ level with the corresponding ratio of mean angular diameter distances 1.63.
We conclude that scattered light from the central source is not the origin of our detection.

Having considered and dismissed a range of alternate explanations for the result shown in Figure \ref{Figure:logFit}, we present quantitative measurements of the H$\alpha$ + [N{\small II}] fluxes as a function of $r_p$ in Table \ref{table:signalstrength}. We present both the directly measured signal in the H$\alpha$ + [N{\small II}] windows and that corrected for the background level, $c$, or in other terms $\Delta f$. From  the corrected values, we have a $3 \sigma$ detection of H$\alpha$ + [N{\small II}] out to $\sim$ 100 kpc. Beyond that radius,  our values are consistent with no flux, but the flux levels predicted by our extrapolated model would not be detectable given the current level of measurement uncertainty in the background. We conclude that we identify ionized gaseous halos that extend to at least $\sim$ 100 kpc around the typical primary galaxy in our sample.

\begin{deluxetable}{ccc}
\tablewidth{0pt}
\tablecaption{Signal Strength In Radial Bins}
\tablehead{
\colhead{$r_p$ range} & \colhead{$\langle \bar{f} \rangle$}  & \colhead{$\Delta f$}\\
\colhead{[Mpc]} & \multicolumn{2}{c}{[$10^{-20} \ {\rm erg} \ {\rm cm}^{-2} \ {\rm s}^{-1} \ {\rm \AA}^{-1}$]}
}
\startdata
$r_p < 0.05$ & 2.25 $\pm$ 0.62 & 2.19 $\pm$ 0.62 \\
$0.05 \le r_p < 0.1$ & 1.16 $\pm$ 0.35 & 1.10 $\pm$ 0.35 \\
$0.1 \le r_p < 0.2$ & 0.06 $\pm$ 0.17 & 0.00 $\pm$ 0.17 \\
$0.2 \le r_p < 0.5$ & 0.14 $\pm$ 0.06 & 0.08 $\pm$ 0.07 \\
$r_p \ge 0.5$ & 0.06 $\pm$ 0.02 & 0.00 $\pm$ 0.04
\enddata
\label{table:signalstrength}
\end{deluxetable}

\subsection{Results Using Subsamples}

\begin{figure*}[!htbp]
\begin{center}
\includegraphics[width = 0.8 \textwidth]{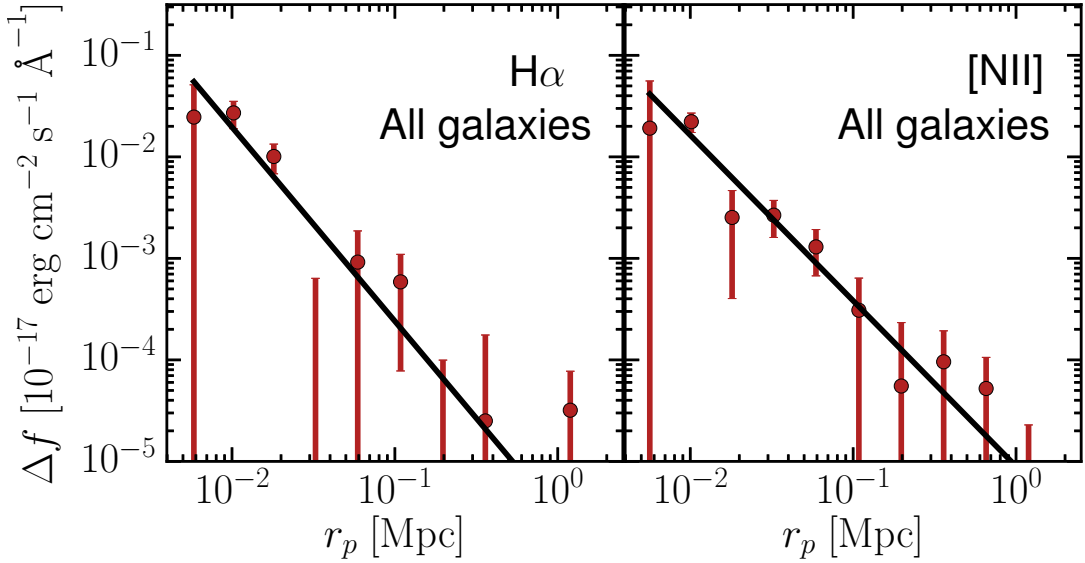}
\end{center}
\caption{The radial profile of  H$\alpha$ flux only (left panel) and [NII] 6548 and 6583 (right panel) for our entire sample. The emission profiles are similar for the different elements.}
\label{Figure:Halpha_fit}
\end{figure*}

The analysis of subsamples allows us to continue to test our results and might provide insights into the physics behind the observations.
So far, we have only considered fits to the combined flux from H$\alpha$ and the two nearby [N{\small II}] lines. We now compare the results using only H$\alpha$ to those using only [N{\small II}]6548 and 6583\AA, shown in Figure \ref{Figure:Halpha_fit}. The quantitative agreement between the two fits confirms the reality of the results, particularly the detection of line emission at $r_p < 100$ kpc. The comparison between the two fits hints that the ratio of [N{\small II}] to H$\alpha$ may rise with radius. This behavior, the apparent hardening of the radiation field with increasing galactocentric radius, has been seen in other studies \citep{miller03} and various causes considered.

\begin{figure*}[!htbp]
\begin{center}
\includegraphics[width = 0.8 \textwidth]{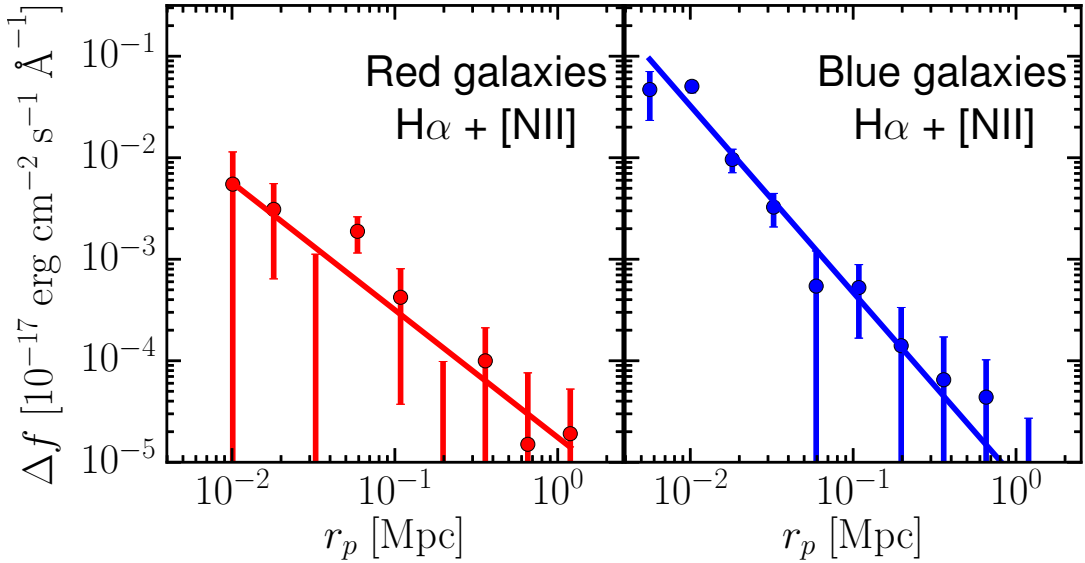}
\end{center}
\caption{The radial profile of H$\alpha$ + [NII] for subsamples divided equally by the color of the primary galaxy (division at $g-r = 0.85)$, with the red subsample in the left panel and the blue subsample in the right panel.
While both subsamples exhibit a rise in the emission flux interior to 100 kpc, the results for the red subsample are weaker and noisier, particularly at small radii. The measurements are consistent beyond 100 kpc.}
\label{Figure:Red_fit}
\end{figure*}

To examine the dependence of the measured halo emission with primary galaxy color, we divide our sample nearly equally into a red class ($g-r \ge 0.85$) and
a blue class ($g - r < 0.85$). This division maps well onto the two color groups identified so clearly in SDSS data \cite[]{2003ApJ...594..186B}.  We present the results from each of the two samples
in Figure \ref{Figure:Red_fit} and Table \ref{table:powerlaw}.

Indeed the blue sample shows stronger central emission that leads to a steeper slope in the emission profile.  Whether the difference extends over all radii is more difficult to say. Nevertheless, the inner difference hints at the importance of escaping radiation or outer disk ionization regions in determining the radial profile of recombination radiation. Perhaps a related effect was identified by \cite{rand96} and \cite{rossa}, where galaxies with higher star formation rates have stronger vertically extended disk emission.
Even so, the
red sample, which consists primarily of low and non-star forming galaxies, also exhibits H$\alpha$ + [N{\small II}] emission at $r_p > 50$ kpc, demonstrating that other sources of ionization are also in play.

Next, we divide the set into two almost equal halves by luminosity or size and present the results in Table \ref{table:powerlaw}. We find very similar radial emission line profiles among samples. There is no strong dependence of halo properties with central galaxy luminosity, over the limited luminosity range of our sample.

We divide the data set into two almost equal halves by the galaxy redshift  and present the results in Table \ref{table:powerlaw}. We find nearly identical radial emission line profiles for the two samples except that the bias level for low redshift galaxies is significantly lower.
We attribute this to the simpler background at bluer wavelengths (Figure \ref{Figure:skybackg}) and conclude that one efficient way to achieve higher sensitivities is to have more lines of sight that probe the nearer primaries in our sample.

\section{Discussion}
\label{sec:discussion}

The interpretation of the emission is complicated by the complex nature of galaxy halos. Although we have excluded lines of sight with significant flux, we could still be contaminated by dwarf galaxies or tidal debris. In the Milky Way halo there is a strong case for significant amounts of ionized gas associated with the Magellanic Stream \citep{fox}.  We have no direct way of assessing this scenario and simple analytic models of halo gas and ionizing radiation fields cannot address it. Full, detailed simulations that follow the baryons and radiation are necessary to examine these questions. Data such as that presented here provides case studies against which to benchmark the simulations.

Our cut, rejecting lines of sight with $|\bar f| > 0.3\times 10^{-17} \ {\rm erg} \ {\rm cm}^{-2} \ {\rm s}^{-1} \ {\rm \AA}^{-1}$, still allows for our measurement to be dominated by a few lines of sight. Our detection of $\langle \bar{f} \rangle \sim 10^{-20} \ {\rm erg} \ {\rm cm}^{-2} \ {\rm s}^{-1} \ {\rm \AA}^{-1}$ suggests that we would only need 1 out of 300 lines of sight to have $\bar f = 0.3\times 10^{-17} \ {\rm erg} \ {\rm cm}^{-2} \ {\rm s}^{-1} \ {\rm \AA}^{-1}$ to obtain the average value that we do. However, the marginal agreement between the mean ($(1.10 \pm 0.35) \times 10^{-20} \ {\rm erg} \ {\rm cm}^{-2} \ {\rm s}^{-1} \ {\rm \AA}^{-1}$) and median ($(0.50 \pm 0.20) \times 10^{-20} \ {\rm erg} \ {\rm cm}^{-2} \ {\rm s}^{-1} \ {\rm \AA}^{-1}$) demonstrates that much of the power of the signal we detect comes from typical lines of sight, while at the same time suggesting that the distribution of $\bar{s}$ is likely to be somewhat asymmetrical. 

Placing this component in context is also critical. Other baryon components have been traced to large radii, for example dust
\citep{zaritsky,nelson,menard}. The most precise measurement of dust out to large radii is that presented by \cite{menard}, who also utilized SDSS data to reach a sensitivity below what was previously available. They find that the surface density of dust is $\propto r_p^{-0.8}$.
The emission line surface brightness we measure is $\propto r_p^{-1.9 \pm 0.4}$ (Table \ref{table:powerlaw}, with an uncertainty here that takes account of the range of variation). If we assume that the gas density follows the dust density, then the volume density of gas and dust is $\propto r^{-1.8}$. Because the emission line flux we see comes from recombination it depends on $n^2$, all other factors being equal, we might expect the volume emission density to be $\propto r^{-3.6}$ and therefore the projected emission density to be $\propto r_p^{-2.6}$, falling roughly in line with what is observed. This argument, or more precisely the uncertainty in the measured radial emission profile, allows for other factors to still play a significant role but it suggests that there are no unexpected large surprises in the dust to gas ratio.

Despite the caveats given regarding the complexity of the actual physical situation, we present a simple calculation in the interest of providing a coarse picture of what may be the situation. The emission rate per cm$^3$ is given by $N_e N_p \alpha_{{\rm H}\alpha}^{\rm eff}$, where $N_e$ and $N_p$ are the electron and proton volume number densities and $\alpha_{{\rm H}\alpha}^{\rm eff}$ is the effective recombination rate (cm$^3$s$^{-1}$). The electron and proton densities are taken to be equal and roughly $10^{-3.8}$  cm$^{-3}$ using $n_{\rm H} = 10^{-4.2}( r_p/{\rm R_{vir}})^{-0.8}$ \citep{werk} and $r_p/{\rm R_{vir}} = 0.3$ for the middle of the 50 to 100 kpc $r_p$ bin. The energy released in one H$\alpha$ photon is $3 \times 10^{-12}$ erg, enabling us to convert the recombination rate into a power. The luminosity distance at a typical redshift in the sample, $z = 0.1$, is 460 Mpc and the angular diameter distance is 380 Mpc. If we assume emission detected in a fiber comes from a 3 arcsec tube on the sky that is 75 kpc in length along the line of sight, we estimate that the measured flux should be $1.6 \times 10^{-7} \alpha_{{\rm H}\alpha}^{\rm eff}$ erg s$^{-1}$ cm$^{-2}$ arcsec$^{-2}$. In these units, our detection in the 50 to 100 kpc bin (Table \ref{table:signalstrength}) corresponds to  $1.9 \times 10^{-20}$ for H$\alpha$ + [N{\small II}] and $1.0 \times 10^{-20}$ for H$\alpha$ only.
For $\alpha_{{\rm H}\alpha}^{\rm eff}$, we adopt the fitting formula presented by \cite{pequignot},
$$ \alpha_{{\rm H}\alpha}^{\rm eff} = 10^{-13} {{2.274\  {\rm T}^{-0.659}}\over{1 + 1.939  \ {\rm T}^{0.574}}},$$
where T is the temperature in units of $10^4$ K. Solving for T, we find that our observed flux, in combination with the density profile from \cite{werk}, result in an estimated temperature of $\sim$12,000 K.

Despite the numerous simplifying assumptions inherent to this temperature estimate, it provides reassurance that the detected signal is physically plausible. Independent temperature estimates for the ``cool" halo gas, in contrast to the million degree coronal gas, based on the detailed analysis of numerous ionized metals, also place the temperature of this component at about 10$^4$ K. In reverse, our observation plus an adopted temperature of $\sim 10^4$ K provides support for the density profile inferred by \cite{werk}. The confirmation of the $n_{\rm H}$ profile further supports their contention that significant mass, as much as 25\% of the total baryonic content of galaxies, resides in this cool component.Simulations also predict that a large fraction of the halo gas resides in a 10$^4$ K phase \citep{keres}. 

\section{Conclusions}
\label{sec:summary}

Our principal result is the first detection of hydrogen recombination radiation from the outer halos of normal galaxies. The presence of gaseous halos had long been inferred from the incidence of metal ion absorption lines seen against the spectra of luminous sources, mostly QSOs, and models demonstrated that this gas was overwhelmingly ionized \citep{putman}. However, searches for hydrogen recombination radiation produced mostly limits except toward individual, H{\small I} clouds, particularly in the Milky Way halo \citep{weiner,fox}.

Using over 7 million spectra that intersect the halos of nearly half a million nearby galaxies, we are able to reach the level of sensitivity required to observe the H$\alpha$ emission up to 100 kpc (Figure \ref{Figure:logFit}). The mean value of the emission from this sample at projected radii between 50 and 100 kpc is $(1.10 \pm 0.35) \times 10^{-20}$
erg s$^{-1}$ cm$^{-2}$ ${\rm \AA}^{-1}$. The units of our measurement are non-standard because we integrate over a fiber aperture but not over the emission line itself. Converting, making some basic assumptions, to other representations our measurement corresponds to $1.87 \times 10^{-20}$ erg cm$^{-2}$ s$^{-1}$ arcsec$^{-2}$ or 0.0033 Rayleigh.

Various tests support the reality of the detection. Our ability to trace this component to larger radii is limited by our ability to set the continuum level.  The uncertainty in our determination of this bias level for our final set of measurements is $\sim 3 \times 10^{-22}$ erg s$^{-1}$ cm$^{-2}$ ${\rm \AA}^{-1}$. We obtain consistent detections for H$\alpha$ and [N{\small II}] independently, for blue and red galaxies, for bright and faint galaxies (within our limited range of luminosities), for small and large galaxies (within our limited range of sizes), and for nearby and distant galaxies (within our limited redshift window). We find that a null test constructed by shuffling the redshifts of our target galaxies does not result in a detection and that we recover an injected line at the flux level of interest. Finally, we combine the detection with a published measurement of the hydrogen number density at these radii to derive a temperature, 12,000 K, that is wholly consistent with both empirical measurements of the temperature based on the metal ions \citep{werk} and numerical simulations the investigate the accretion of gas onto galaxies \citep{keres}.

Our detection of recombination emission from the dominant component of the cool gas in galaxy halos provides a benchmark for future studies that will examine it in more detail and opens a new avenue for the exploration of galaxy evolution.

\section{Acknowledgments}
DZ acknowledges financial support from NASA ADAP NNX12AE27G and NSF grant AST-1311326. The authors gratefully acknowledge the SDSS III team for providing a valuable resource to the community.

Funding for SDSS-III has been provided by the Alfred P. Sloan Foundation, the Participating Institutions, the National Science Foundation, and the U.S. Department of Energy Office of Science. The SDSS-III web site is http://www.sdss3.org/.

SDSS-III is managed by the Astrophysical Research Consortium for the Participating Institutions of the SDSS-III Collaboration including the University of Arizona, the Brazilian Participation Group, Brookhaven National Laboratory, Carnegie Mellon University, University of Florida, the French Participation Group, the German Participation Group, Harvard University, the Instituto de Astrofisica de Canarias, the Michigan State/Notre Dame/JINA Participation Group, Johns Hopkins University, Lawrence Berkeley National Laboratory, Max Planck Institute for Astrophysics, Max Planck Institute for Extraterrestrial Physics, New Mexico State University, New York University, Ohio State University, Pennsylvania State University, University of Portsmouth, Princeton University, the Spanish Participation Group, University of Tokyo, University of Utah, Vanderbilt University, University of Virginia, University of Washington, and Yale University.

\bibliography{bibliography}

\end{document}